# Tiny Chiplets Enabled by Packaging Scaling: Opportunities in ESD Protection and Signal Integrity

Emad Haque[1], Pragnya Sudershan Nalla[2], Jeff Zhang[1], Sachin S. Sapatnekar[2],
Chaitali Chakrabarti[1] and Yu Cao*[2]

[1] Arizona State University, School of Electrical, Computer and Energy Engineering, Tempe, AZ, USA
[2] University of Minnesota, Department of Electrical and Computer Engineering, Minneapolis, MN, USA
* Email: yucao@umn.edu

*Abstract*—The scaling of advanced packaging technologies provides abundant interconnection resources for 2.5D/3D heterogeneous integration (HI), thereby enabling the construction of larger-scale VLSI systems with higher energy efficiency in data movement. However, conventional input/output (I/O) circuitry, including electrostatic discharge (ESD) protection and signaling, introduces significant area overhead. Prior studies have identified this overhead as a major constraint in reducing chiplet size below 100 mm$^2$. In this study, we revisit reliability requirements from the perspective of chiplet interface design. Through parasitic extraction and simulation program with integrated circuit emphasis (SPICE) simulations, we demonstrate that ESD protection and inter-chiplet signaling can be substantially simplified in future 2.5D/3D packaging technologies. Such simplification, in turn, paves the road for further chiplet miniaturization and improves the composability and reusability of tiny chiplets.

## I. Introduction

Heterogeneous integration (HI) systems, which leverage 2.5D and 3D packaging technologies to integrate multiple chiplets on an advanced packaging substrate, have emerged as a key enabler in modern very large-scale integration (VLSI) design [1, 2]. These systems offer high flexibility, scalability, and high energy efficiency, particularly for data-intensive workloads such as high-performance computing, autonomous vehicles and AI tasks. By disaggregating a large monolithic design into smaller chiplets and interconnecting them through 2.5D and 3D integration, these systems effectively reduce design and fabrication costs, improve overall yield, provide higher bandwidth and lower energy consumption in data movement, and achieve better system reconfigurability [3, 4].

However, previous studies have identified the I/O interface of chiplets, such as electrostatic discharge (ESD) protection, clock and data synchronization, and related area cost as a key limitation in scaling chiplet sizes below 100 mm$^2$ [5]. For instance, an implementation of the Advanced Interface Bus (AIB) [6], a widely adopted I/O module for 2.5D integration, occupies several mm$^2$ at 22nm, larger than many design IP blocks (such as CPUs, DSPs, FFT accelerators, and systolic arrays) [7]. These overheads restrict the reusability and composability of chiplets in heterogeneous integration [7].

In this study, we revisit the reliability requirements of chiplet interfaces, with a focus on advanced packaging technologies. As packaging continues to scale, with finer pitch, shorter inter-chiplet spacing, and lower electrical parasitics, we explore how these advancements can significantly reduce the overhead of ESD protection and inter-chiplet signaling, thereby eliminating conventional I/O bottlenecks and enabling future scaling of chiplet sizes (i.e., tiny chiplets).

## II. Scaling of 2.5D/3D Packaging

A heterogeneous system typically comprises three primary components [1, 2]: the chiplets, which serve as the functional units for computing, memory, control and other tasks; the interconnects, either horizontal or vertical, that connect the chiplets together to form a complete system; and the substrate, which can be silicon-, organic- or glass-based, providing the foundation for hosting both the interconnects and chiplets. Figure 1 illustrates such a heterogeneous system with multiple stacks of 2.5D and 3D chiplets on a common substrate, highlighting key features and parameters of wires between chiplets. Figure 2 presents a more detailed view of an individual 2.5D chiplet, with μbump arrays on each side for delivering signals, clocks, and power supply.

As packaging technologies advance, several physical features of HI systems continue to scale down. Table I outlines important geometric parameters related to the chiplet size and interconnect dimensions. Based on HI roadmaps [1, 2], Tables II and III summarize the scaling trends of μbumps and hybrid bonds. These two structures form the critical interface between chiplets and the substrate or between chiplets themselves. Their dimensions directly influence the electrical properties relevant to ESD and signal integrity analysis.

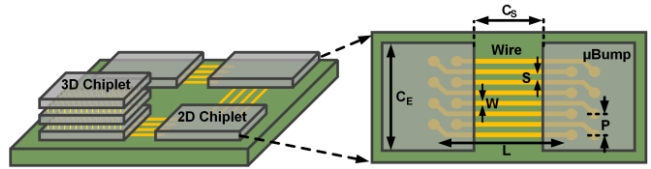

Fig. 1. Overview of a 2.5D/3D heterogeneous system.

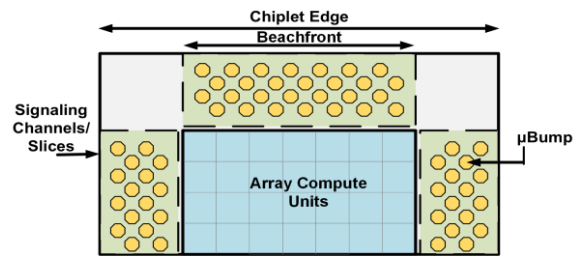

Fig. 2. Top view of a 2.5D chiplet with compute units and μbump arrays as the inter-chiplet interface [1, 2].

TABLE I. Physical Geometry in a HI System.

| Parameter | Description | Parameter | Description |
|---|---|---|---|
| $C_E$ | Chiplet edge length | S | Wire spacing |
| $C_S$ | Chiplet spacing | H | Wire height |
| L | Wire length | T | Wire thickness |
| W | Wire width | P | μbump pitch |

TABLE II. SCALING OF μBUMP TECHNOLOGIES [1, 2].

| Generation | 0 | 1 | 2 | 3 | 4 | 5 |
|---|---|---|---|---|---|---|
| L (mm) | 4 | 2 | 1.6 | 0.9 | 0.4 | 0.15 |
| W = S (μm) | 2.5 | 2 | 1.5 | 1 | 0.5 | 0.25 |
| T = H (μm) | 5 | 4 | 3 | 2 | 1 | 0.5 |
| P (μm) | 70 | 55 | 40 | 30 | 20 | 10 |

TABLE III. SCALING OF HYBRID BONDING TECHNOLOGIES [1, 2].

| Generation | 0 | 1 | 2 | 3 | 4 |
|---|---|---|---|---|---|
| L (μm) | 150 | 100 | 75 | 50 | 25 |
| W = S (μm) | 0.25 | 0.20 | 0.15 | 0.10 | 0.05 |
| T = H (μm) | 0.5 | 0.4 | 0.3 | 0.2 | 0.1 |
| P (μm) | 10 | 5 | 2.5 | 1 | 0.5 |

### III. ELECTROSTATIC DISCHARGE PROTECTION

One of the key cost factors in chiplet-based design is electrostatic discharge protection for the I/O interface [5, 8]. ESD poses a critical concern for the reliability of on-chip transistors, and its adverse impact becomes even more pronounced as complementary metal oxide semiconductor (CMOS) technology continues to scale. At the chiplet level, only the interfaces, such as μbumps and hybrid bonds, are exposed to potential ESD from the external environment and therefore, require protection.

ESD protection typically involves large diodes within I/O cells that clamp the ESD voltage below the damage threshold. However, this approach inevitably introduces significant area overhead and additional capacitive load on the signal channel, thereby limiting the usable chip area and reducing the data rate of inter-chiplet communication. Given the controlled clean room environment in advanced packaging, the JEDEC roadmap recommends reducing ESD protection targets from 250V today to 125V for scaled packaging technologies, and even down to as low as 5V for chiplet I/Os using hybrid bonding in 2.5D/3D systems [8].

To meet the JEDEC requirements of ESD protection, it is essential to determine the minimum diode size, which is affected by many factors, such as the parasitic resistance (R), inductance (L), and capacitance (C) along the I/O path. In this study, we use the circuit schematic for ESD validation, as shown in Fig. 3 [9, 10]. We perform SPICE simulations to evaluate the diode size needed for ESD protection under the Charged Device Model (CDM) across generations. This generic schematic includes both the interconnect component and the pad structure, such as μbumps and hybrid bonds used in 2.5D/3D integration. Based on the dimensions listed in Table II, we adopt compact models to calculate the corresponding RLC parameters for each generation [11]. As an example, Table IV summarizes the parameters for μbumps.

Using the values in Table IV, we simulate the gate voltage waveforms with μbumps to search for the required diode size. It is determined by applying the target voltage to the package capacitance and checking whether the gate voltage is below the gate oxide breakdown voltage. Table V lists the minimum size necessary to keep the gate voltage below the gate-oxide breakdown voltage (3.8V for the 28 nm process node). Figure 4 presents the gate voltage waveforms at the minimum diode size, confirming that all voltages remain below the threshold. As observed in Table V, the diode area of μbumps decreases across generations, because dimension leads to a lower impedance path, thereby improving the protection for downstream transistors. Nevertheless, for μbump technologies, a significant area must still be allocated for the diode to drive the interconnect in 2.5D integration.

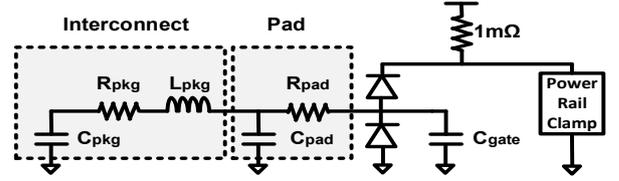

Fig. 3. SPICE simulation model to evaluate the required ESD diode size for CDM protection [9, 10].

TABLE IV. PARAMETERS FOR μBUMPS IN SPICE SIMULATIONS.

| Parameter | $C_{pkg}$ (fF) | $R_{pkg}$ (Ω) | $L_{pkg}$ (nH) | $C_{pad}$ (fF) | $R_{pad}$ (mΩ) |
|---|---|---|---|---|---|
| Gen 0 | 1141.04 | 7.040 | 5.978 | 5.911 | 4.574 |
| Gen 1 | 423.11 | 11.00 | 2.801 | 4.645 | 5.822 |
| Gen 2 | 427.89 | 7.333 | 2.262 | 3.378 | 8.004 |
| Gen 3 | 233.26 | 9.899 | 1.242 | 2.533 | 10.673 |
| Gen 4 | 103.67 | 17.599 | 0.542 | 1.689 | 16.009 |
| Gen 5 | 38.87 | 26.400 | 0.195 | 0.844 | 32.018 |

The situation improves substantially for hybrid bonding with shorter 2.5D inter-chiplet spacing. The RLC parameters are calculated from the dimensions in Table III. Due to the much smaller feature sizes here as compared to μbumps, the need for diodes is eliminated. Figure 5 presents the voltage waveforms for a 10V ESD event (above the 5V JEDEC threshold). Even without any ESD protection diodes, the voltage remains below 3.8V. This indicates that 2.5D tiny

TABLE V. REQUIRED ESD DIODE AREAS AT 28NM WITH μBUMPS.

| ESD Target Voltage (V) | Total Required Diode Area (μm²) | | | | | |
|---|---|---|---|---|---|---|
| | Gen 0 | Gen 1 | Gen 2 | Gen 3 | Gen 4 | Gen 5 |
| 10 | 6.46 | 6.11 | 6.38 | 6.15 | 6.15 | 6.02 |
| 30 | 20.4 | 18.2 | 20.1 | 18.7 | 18.7 | 17.9 |
| 50 | 34.3 | 30.5 | 34.2 | 31.2 | 31.2 | 29.8 |
| 125 | 51.8 | 46.6 | 51.6 | 47.9 | 47.9 | 45.4 |

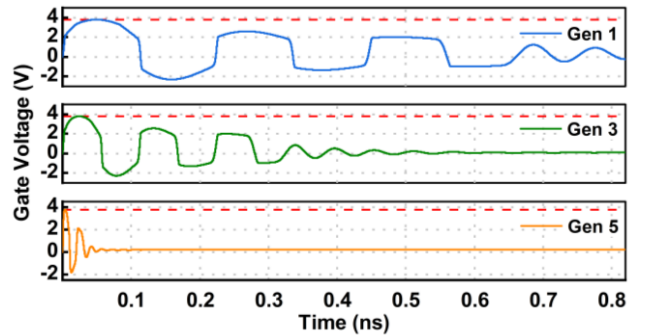

Fig. 4. Simulated gate voltage for a 125V ESD event for μbumps at Gen 1, Gen 3, and Gen 5. Gate-oxide breakdown voltage is marked in red.

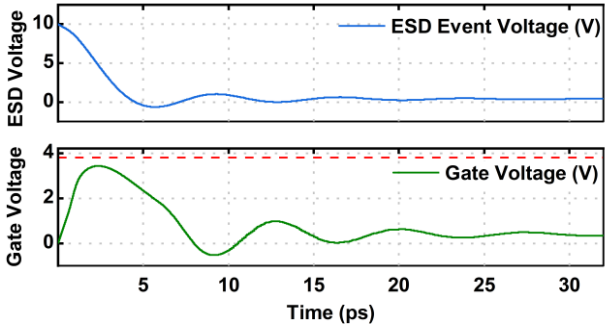

Fig. 5. ESD event voltage and gate voltage for hybrid bonding at Generation 4, without using any ESD diode for protection. Gate-oxide breakdown voltage is marked in red.

chiplets with hybrid bonding can eliminate the expensive on-chiplet ESD protection.

## IV. INTER-CHIPLET SIGNALING

For 2.5D inter-chiplet communication, multiple I/O protocol standards have been proposed, including AIB [6], Universal Chiplet Interconnect Express (UCIe) [12] and Bunch of Wires (BoW) [13]. These protocols typically need to drive interconnects spanning several millimeters, as shown in Table II. As a result, they are often adapted from conventional high-speed I/O designs, such as double data rate (DDR) and peripheral component interconnect express (PCIe), and then customized for the specific packaging wire load. Like DDR and PCIe, these modules are implemented as IP blocks integrated at the chiplet edges, but they still involve considerable design complexity and area cost.

In this study, we explore direct signaling link (DSL), a much simpler I/O approach that resembles multiple parallel data buses within a chiplet. It is inspired by the observation that inter-chiplet spacing is approaching the typical length of on-chip interconnects. With simplified transceivers and receivers, essentially on-chip buffers, DSL offers a lightweight signaling structure that is particularly appealing for tiny chiplets. Figure 6 presents the DSL schematic used in our SPICE simulations. The CMOS driver is determined by the required bandwidth and channel parasitics to ensure signal integrity. The receiver is the same size as the driver.

To assess the signal integrity of DSL in inter-chiplet communication, we examine eye diagrams under various configurations for a fixed driver size. Figure 7 presents the eye diagrams at 1GHz for four generations of μbumps. As packaging technology scales, our analysis shows continuous improvement in signal quality, making DSL increasingly viable. Figure 8 further studies eye diagrams for Generation 5 with varying channel length. As spacing decreases, DSL demonstrates better quality in signal transmission. Therefore, in advanced packaging where inter-chiplet distances for adjacent chiplets decrease to hundreds of micrometers, DSL emerges as a suitable I/O solution for tiny chiplets. However, for longer channel lengths in the order of millimeters, special protocols, such as AIB or UCIe, remain necessary. Note that when DSL is used for signaling across a short distance, clock synchronization can also be relaxed by leveraging globally asynchronous locally synchronous (GALS) architectures or similar techniques.

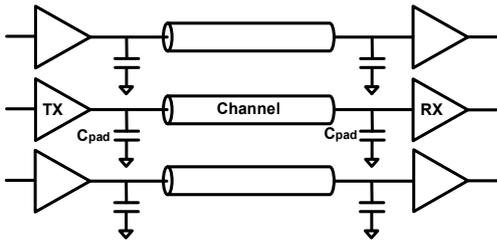

Fig. 6. SPICE models to simulate Direct Signaling Links, with three parallel channels shown as an example. The two outer channels act as aggressors to evaluate crosstalk noise. (TX: transceiver; RX: receiver)

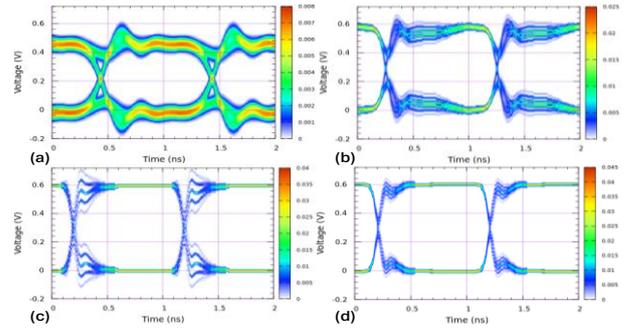

Fig. 7. Eye diagrams for DSL for a fixed driver size, for (a) Gen 0, (b) Gen 1, (c) Gen 3, (d) Gen 5.

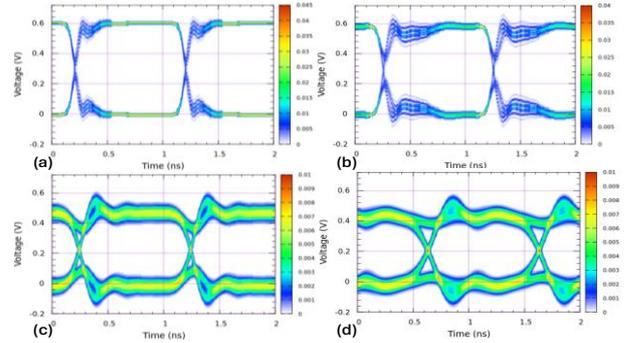

Fig. 8. Eye diagrams of DSL for packaging Gen 5, with channel length of (a) 150 μm, (b) 750 μm, (c) 2 mm, and (d) 4 mm.

## V. IMPACT ON TINY CHIPLETS

As highlighted in previous studies, the cost of ESD and other circuitry in I/O cells has long been the key barrier to chiplet scaling [1, 2, 5]. This limitation further restricts the flexibility and composable of chiplets, as reusable, low-cost hardened IPs across a broad range of applications.

Based on our simulations in Sections III and IV, we demonstrate that such I/O costs are expected to decrease significantly as packaging technologies continue to scale down, with shorter and narrower wires and smaller interface pads. For μbumps, Fig. 9(a) summarizes the area of ESD protection diode per pad. Use of hybrid bonding along with new I/O structures and cleaner assembly environments, helps eliminate the need for on-chip ESD protection, dramatically simplifying I/O design and planning.

Furthermore, Fig. 9(b) compares the total area of I/O bump arrays for both AIB and the proposed DSL scheme. At shorter distances, DSL not only ensures the integrity in inter-chiplet signaling, but also reduces the area cost by more than 8× compared to AIB. At longer distances, the special I/O protocols are necessary for signal integrity. The AIB I/O array area saturates above a 4 mm edge length due to its architectural limit of 24 channels per column. With a fixed bump pitch (10 μm in this study), this channel limit becomes the dominant constraint, and increasing the edge length no longer improves I/O density. Overall, DSL delivers a compact and cost-effective I/O solution than current 2.5D I/O modules for future tiny chiplets.

2.5D/3D advanced packaging offers a key advantage over monolithic design by improving I/O data rates. This improvement stems from the increased number of pads and the use of narrower wires to boost channel density. For both AIB and DSL, the implementations are aligned along the chiplet edge, which inherently limits the number of available pads and the achievable data bandwidth. In this scenario, the I/O bandwidth scales linearly with the chiplet edge length, while the computing capability and its associated data demands increase quadratically with chiplet size. This mismatch implies that as chiplet size increases, performance will eventually be limited by I/O bandwidth.

To illustrate this, we set up the general chiplet structure, following Fig. 2, consisting of a systolic array of multiply-accumulate (MAC) units for compute and μbump arrays for inter-chiplet signaling [14]. Based on the number of available channels and compute units, we derive both the maximum supported bandwidth, as well as the data rate required to fully utilize the compute resources. Figure 10(a) shows that for an older technology node, both AIB and DSL can meet the data demands of compute units up to a chiplet size of approximately 10 mm × 10 mm (i.e., 100 mm$^2$). However, for more advanced nodes in Fig. 10(b), AIB increasingly struggles to meet the data requirements due to its larger area overhead. In contrast, the more compact design of DSL continues to support higher data bandwidths, enabling tiny chiplets down to 2 mm × 2 mm (i.e., 4 mm$^2$). This confirms our earlier observations on the benefits of chiplet scaling.

## VI. CONCLUSION

With lower packaging parasitics, the need of ESD protection is significantly reduced or even eliminated in future HI systems. High-speed inter-chiplet signaling is further supported by simpler circuits, such as direct signaling links, than 2.5D I/O modules today (AIB, UCIe and BoW). Such a trend enables the scaling of tiny chiplets down to 4 mm$^2$, leading to better composability and IP reusability.

## ACKNOWLEDGMENT

This work is supported in part by COCOSYS, one of six centers in JUMP 2.0, a Semiconductor Research Corporation (SRC) program sponsored by DARPA. This work is also partially supported by National Science Foundation under grant CCF-2403408 and CCF-2403409.

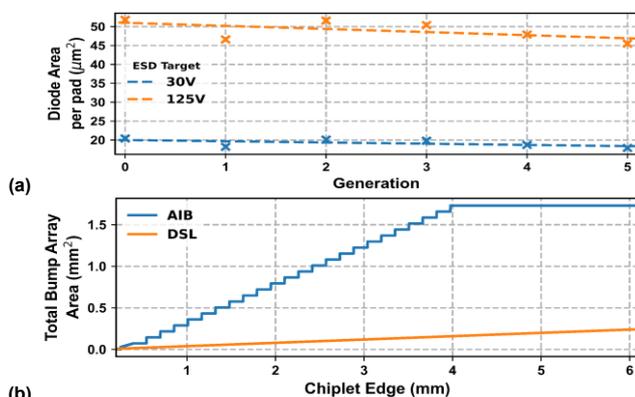

Fig. 9. Area overhead in (a) ESD protection per pad for μbumps across generations; and (b) total bump array of AIB and DSL vs. chiplet edge size.

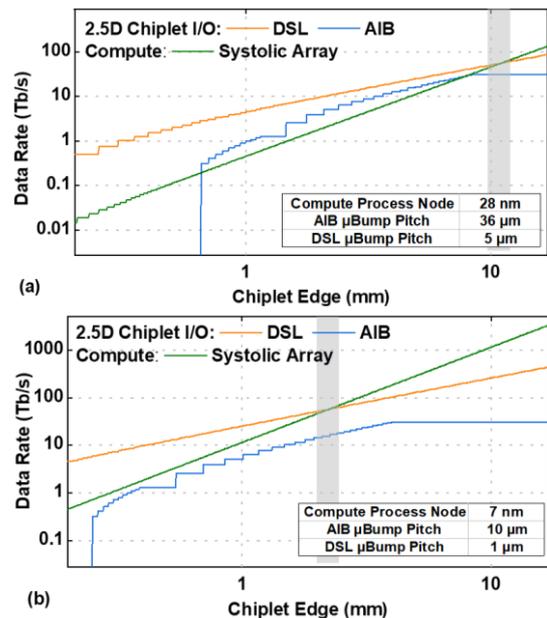

Fig. 10. Effect of chiplet size on the maximum achievable data rate for AIB and DSL interfaces per chiplet. DSL is able to support the high data rate required for smaller chiplet sizes (< 4 mm$^2$) through packaging scaling. Array compute is the theoretical maximum data rate when all array units in the chiplet transfer data in parallel.